\documentclass[conference,9pt,dvipsnames]{IEEEtran}

\AtBeginDocument{%
  \providecommand\BibTeX{{%
    \normalfont B\kern-0.5em{\scshape i\kern-0.25em b}\kern-0.8em\TeX}}}
\usepackage{cite}

\usepackage{url}
\usepackage{enumitem}

\usepackage[utf8]{inputenc} 
\usepackage[T1]{fontenc}    
\usepackage{url}            
\usepackage{booktabs}       
\usepackage{amsfonts}       
\usepackage{amsmath}
\usepackage{nicefrac}       
\usepackage{microtype}      
\usepackage{xcolor}         
\usepackage{colortbl}
\usepackage{graphicx}
\usepackage{svg}
\usepackage{multirow}
\usepackage{wrapfig}
\usepackage{threeparttable}
\usepackage{tikz}
\usepackage{makecell}
\usepackage{tablefootnote}
\usepackage[hidelinks]{hyperref}

\usepackage{color}
\usepackage{xspace}
\usepackage{caption}
\usepackage{adjustbox}

\usepackage{subfigure}

\usepackage{amsmath}
\usepackage{amssymb}
\usepackage{mathtools}
\usepackage{amsthm}
\usepackage{hyperref}
\usepackage{adjustbox}
\usepackage{xcolor}
\theoremstyle{plain}
\newtheorem{theorem}{Theorem}[section]

\newtheorem{lemma}[theorem]{Lemma}

\theoremstyle{definition}
\newtheorem{definition}[theorem]{Definition}

\theoremstyle{remark}

\begin{document}
\pagestyle{plain}
\title{\name: Instruction Alignment in Large Language Models for Chip Design via Geodesic Interpolation}


\author{
    \IEEEauthorblockN{Chenhui Deng}
    \IEEEauthorblockA{ 
        NVIDIA \\
        \centering \, cdeng@nvidia.com
    }
    \and
    \IEEEauthorblockN{Yunsheng Bai}
    \IEEEauthorblockA{
        NVIDIA \\
        \centering \, yunshengb@nvidia.com
    }
    \and
    \IEEEauthorblockN{Haoxing Ren}
    \IEEEauthorblockA{
        NVIDIA \\
        \centering \, haoxingr@nvidia.com
    }
}



\newcommand{\name}{ChipAlign\xspace}

\newcommand{\red}[1]{\textcolor{red}{#1}}
\newcommand{\cd}[1]{\textcolor{violet}{[chenhui: #1]}}
\newcommand{\yba}[1]{\textcolor{blue}{[Yunsheng: #1]}}


\maketitle

\begin{abstract}
Recent advancements in large language models (LLMs) have expanded their application across various domains, including chip design, where domain-adapted chip models like ChipNeMo have emerged. However, these models often struggle with instruction alignment, a crucial capability for LLMs that involves following explicit human directives. This limitation impedes the practical application of chip LLMs, including serving as assistant chatbots for hardware design engineers. In this work, we introduce ChipAlign, a novel approach that utilizes a \textit{training-free} model merging strategy, combining the strengths of a general instruction-aligned LLM with a chip-specific LLM. By considering the underlying manifold in the weight space, ChipAlign employs geodesic interpolation to effectively fuse the weights of input LLMs, producing a merged model that inherits strong instruction alignment and chip expertise from the respective instruction and chip LLMs. Our results demonstrate that ChipAlign significantly enhances instruction-following capabilities of existing chip LLMs, achieving up to a $26.6\%$ improvement on the IFEval benchmark, while maintaining comparable expertise in the chip domain. This improvement in instruction alignment also translates to notable gains in instruction-involved QA tasks, delivering performance enhancements of $3.9\%$ on the OpenROAD QA benchmark and $8.25\%$ on production-level chip QA benchmarks, surpassing state-of-the-art baselines. 
\end{abstract}

\section{Introduction}

Recent years have seen remarkable breakthroughs in large language models (LLMs) across a wide range of applications, including text summarization, machine translation, and conversational AI~\cite{zhao2023survey}. Models like GPT~\cite{achiam2023gpt}, Gemini~\cite{team2023gemini}, Claude~\cite{TheC3}, and LLaMA~\cite{dubey2024llama} series have transformed numerous industries by automating complex tasks, enhancing decision-making processes, and enabling creative problem-solving that traditionally required human expertise. Alongside this success, there is a growing trend of adapting LLMs to specific domains to meet specialized needs. Domain-adapted LLMs have been developed for fields such as healthcare~\cite{wu2024pmc}, finance~\cite{wu2023bloomberggpt}, law~\cite{cui2023chatlaw}, and climate~\cite{vaghefi2023chatclimate}, where nuanced understanding and specialized knowledge are essential for enhancing model performance within these domains.

In the realm of chip design, ChipNeMo stands out as a prominent example of domain-adapted LLMs~\cite{liu2023chipnemo}. 
Built on the LLaMA2-70B foundation model, ChipNeMo leverages domain-adaptive pretraining (DAPT) and finetuning (DAFT) to imbue the model with specialized knowledge in circuits, bugs, and electronic design automation (EDA) scripts.
Following ChipNeMo, several customized LLMs have recently been developed for EDA tasks. He et al. introduced AutoMage, an LLM finetuned on LLaMA2 to specialize in EDA tool utilization~\cite{wu2024chateda}. Subsequently, Sharma et al.~\cite{sharma2024openroad} and Pu et al~\cite{pu2024customized}. developed LLMs tailored for QA and script generation tasks from OpenROAD.

\begin{figure}[t!]
\begin{center}
	\includegraphics[width=\columnwidth]{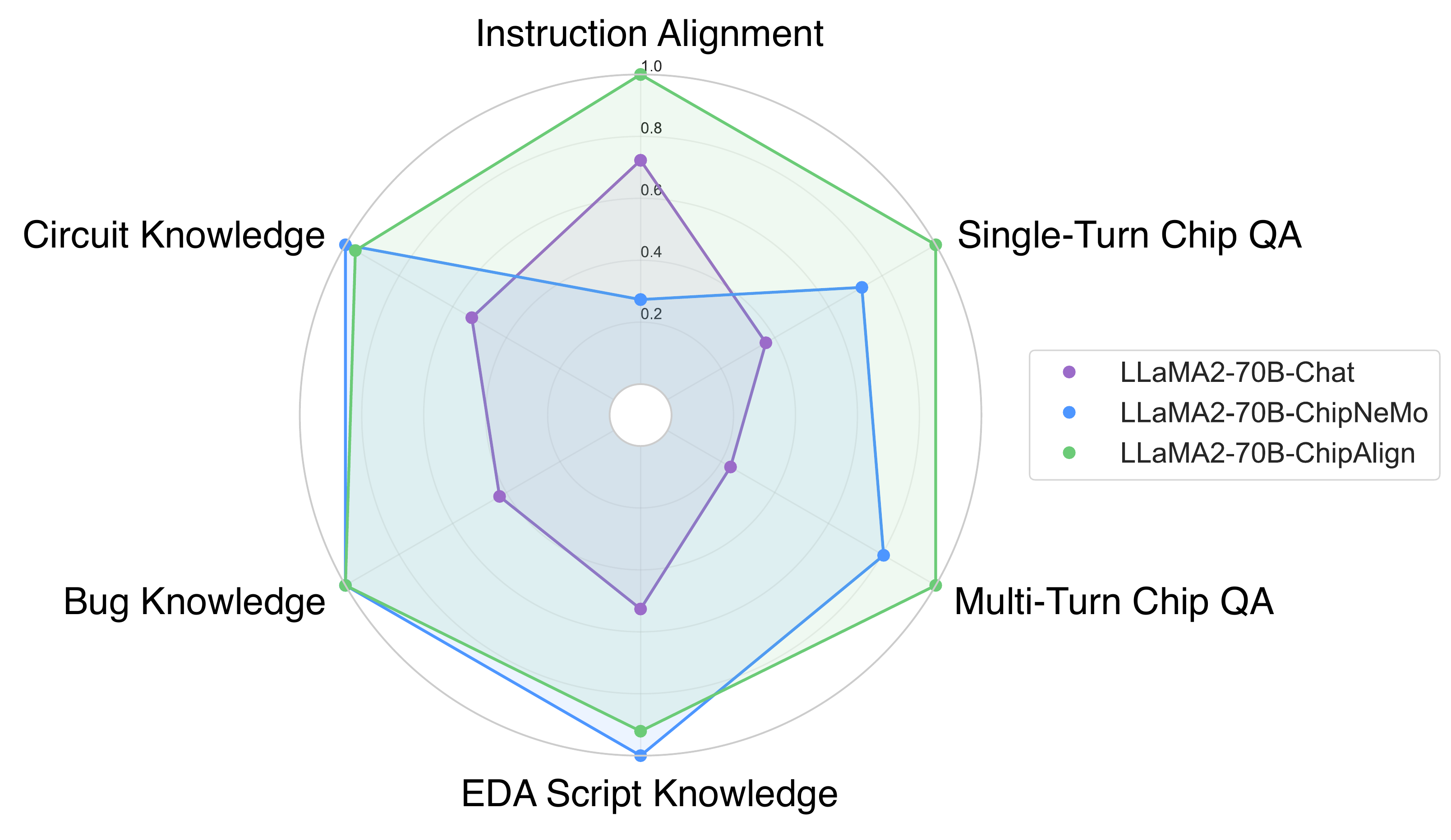}
    \vspace{-15pt}
	\caption{An overview of the capabilities of LLaMA2-70B variants: Chat, ChipNeMo, and ChipAlign on instruction alignment and chip domain benchmarks --- Scores are normalized to $[0,1]$ as per \cite{li2024baichuan}, with points closer to the outermost circle indicating better performance.
    } 
     \label{figure:radar}
\end{center}
\vspace{-20pt}
\end{figure}

However, these tailored LLMs typically exhibit 
diminished instruction alignment,
a fundamental capability of general-purpose chat LLMs to follow human instructions, as demonstrated in Figure \ref{figure:radar}. 
This decline in instruction alignment limits the practical usability of chip LLMs, as they may struggle to respond effectively to user-directed commands, making them less versatile and reliable in real-world applications. 
For instance, when serving as single or multi-turn chatbots for hardware design engineers,
it is critical that the chip LLMs not only possess deep hardware knowledge but also strictly adhere to engineer instructions, such as 
``Please answer questions exclusively based on the provided context'' or ``Please provide a detailed and rigorous solution''.
Unfortunately, prior chip LLMs often compromise this ability after going through DAPT or DAFT, leading to less satisfactory responses as indicated in Figures \ref{figure:prompt} and \ref{figure:multiturn}.

\begin{figure}[t!]
\begin{center}
	\includegraphics[width=\columnwidth]{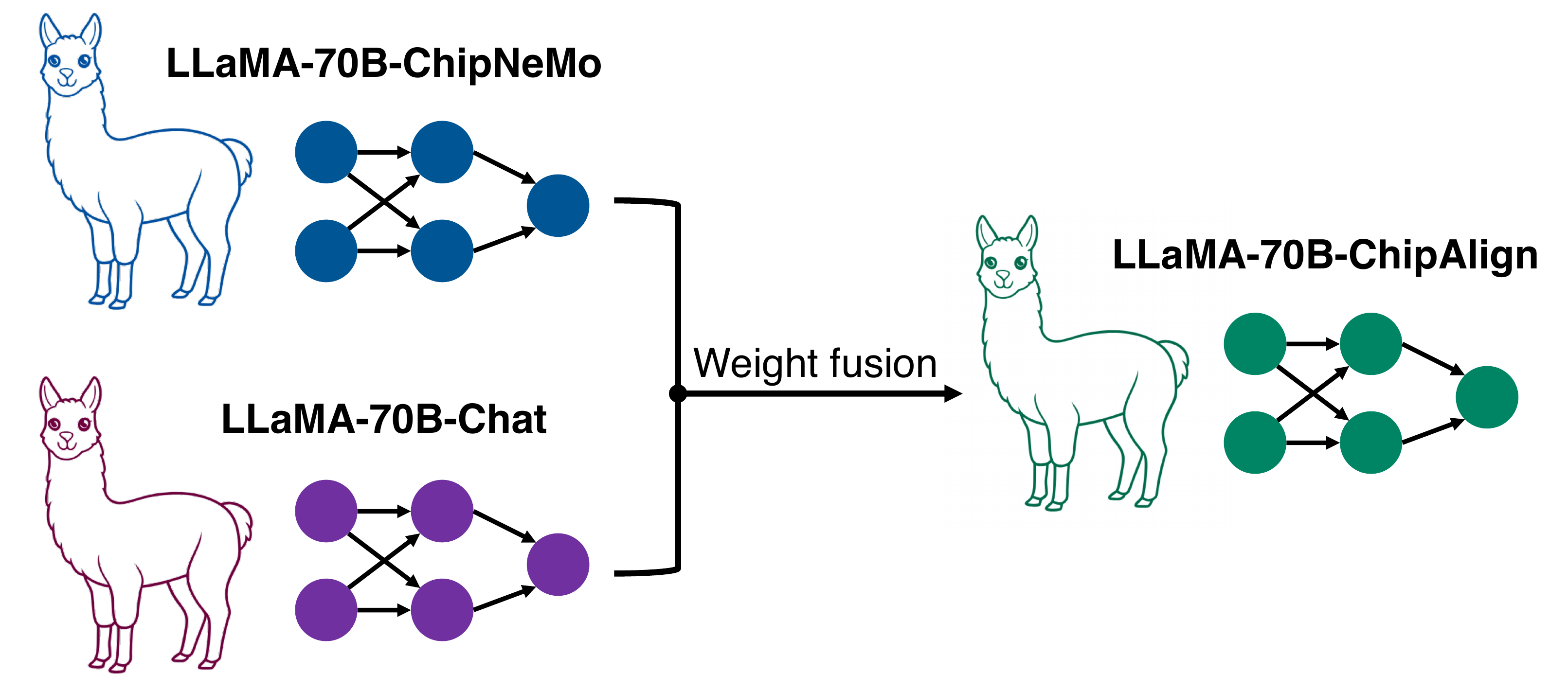}
    \vspace{-15pt}
	\caption{An illustration of model merging.} 
    \vspace{-15pt}
     \label{figure:merge}
\end{center}

\end{figure}
To overcome the aforementioned challenge, a viable solution is multi-task learning, training an LLM on both chip domain-specific and instruction-following data. This approach equips the model with dual capabilities, combining chip design expertise with a strong instruction-following capability. Nonetheless, a significant obstacle to multi-task learning is the proprietary nature of high-quality instruction datasets, such as those used even for open-sourced models like the LLaMA series~\cite{dubey2024llama}, rendering it difficult to obtain a well-aligned model. 

In this work, we address the instruction alignment challenge from an orthogonal perspective. Instead of training on instruction data, we leverage a \textit{training-free} approach based on model merging, which fuses the weights of LLMs specialized in different tasks to produce a unified model that excels across them, as shown in Figures \ref{figure:merge}. To this end, we introduce \name, which enhances the instruction alignment of a chip LLM by merging its weights with those of a well-aligned general instruction model that is publicly available (e.g., LLaMA2-Chat or LLaMA3-Instruct). 
More concretely, we treat the weights of both instruction and chip LLMs as two points on a Riemannian manifold, a curved geometric space that allows for measuring distances and angles~\cite{amari1992information}. 
To construct a merged model that embodies both strong instruction alignment and chip design expertise, it should ideally reside near both input models on this manifold. This requirement implies positioning the model along the geodesic—the shortest path between the two points on the manifold. Accordingly, \name employs geodesic interpolation to generate a new model that effectively combines the strengths of both input models. 

To demonstrate the viability and flexibility of \name, we conduct extensive experiments across several benchmarks, including the instruction-following evaluation (IFEval) benchmark~\cite{zhou2023instruction} and various chip domain benchmarks such as the open-source OpenROAD QA benchmark~\cite{pu2024customized} and proprietary production-level chip QA benchmarks. We utilize LLMs ranging from 8B to 70B parameters. Our results indicate that \name significantly enhances the instruction-following capabilities of chip LLMs, thereby improving their performance on practical chip QA benchmarks that involve specific instructions.
We summarize our main technical contributions as follows:

\noindent $\bullet$ To our knowledge, we are the first to apply a model merging approach to domain-adapted LLMs in chip design. This is achieved by fusing the weights of a chip LLM with those of an instruction-aligned LLM, without requiring additional training.

\noindent $\bullet$ By considering the geometric properties of the weight space in LLMs, \name utilizes geodesic interpolation to produce a merged model. 
This facilitates a smoother fusion of input LLM weights, leading to consistent enhancements in performance over previous model merging methods across different LLM backbones.

\noindent $\bullet$ Owing to the geometric-aware merging scheme in \name, the merged model inherits both strong instruction alignment and chip expertise from the input LLMs. Compared to the state-of-the-art baseline ChipNeMo, \name achieves a significant enhancement in instruction alignment, showing a $26.6\%$ improvement on the IFEval benchmark while maintaining comparable domain knowledge.

\noindent $\bullet$ Thanks to the enhanced instruction alignment, \name outperforms state-of-the-art chip LLMs in instruction-involved QA tasks, achieving improvements of $3.9\%$ on the OpenROAD QA benchmark and $8.25\%$ on production-level chip QA benchmarks.

\section{Background}
\subsection{Domain-Adapted LLMs for Chip Design}
\label{da_llm}
In chip design, the need for domain-adapted LLMs has driven the development of several models tailored to hardware-related tasks. Liu et al. developed ChipNeMo starting with DAPT on 24 billion tokens drawn from chip design documents and code, using the LLaMA2-70B foundation model as a base. This pretraining phase employs the standard autoregressive language modeling objective to tailor the model to domain-specific data. Subsequently, the model underwent DAFT on approximately 57,000 samples, incorporating both domain-specific instructional data and open-source chat data from OASST~\cite{kopf2024openassistant}. 
Through DAPT and DAFT, ChipNeMo acquired specialized knowledge in the chip domain, leading to promising outcomes in various hardware design applications~\cite{liu2023chipnemo}. Later, He et al. developed AutoMage, an LLM finetuned on LLaMA2 for EDA tool usage~\cite{wu2024chateda}, which led to the creation of ChatEDA, an autonomous agent customized for EDA design flow.
More recently, Sharma et al.~\cite{sharma2024openroad} and Pu et al.~\cite{pu2024customized} have tailored LLMs for OpenROAD script generation and QA tasks, covering a broad spectrum of queries related to command usage, VLSI flow, installation guides, and GUI usage.

\subsection{Instruction Alignment in Chip Design LLMs}
\label{instruct_align}
DAPT and DAFT often drastically change the weights of LLMs to emphasize domain knowledge, 
resulting in a loss of the instruction-following capabilities originally present in general-purpose LLMs~\cite{ghosh2024closer}. However, instruction alignment is crucial for real-world applications, such as a chatbot assistant for chip designers. In such settings, designers may seek guidance on design methodologies, troubleshooting steps, or explanations of specific design concepts, often phrased as direct instructions. Additionally, they may instruct the chatbot to respond solely based on a given context, which ensures the answer is grounded in relevant and context-specific information, as shown in Figures \ref{figure:prompt} and \ref{figure:multiturn}. 
Hence, the ability to understand and respond appropriately to these instructions is vital for the practical usability of a chip LLM, making instruction alignment an essential feature.

A straightforward approach to enhance the instruction alignment of chip LLMs involves multi-task learning, which simultaneously trains a model on chip domain-specific data and instruction-following data to effectively integrate both sets of capabilities.
However, access to high-quality instruction data is limited, as datasets used by advanced models like GPT-4 and the LLaMA series remain proprietary. 
While open-source instruction datasets are valuable~\cite{kopf2024openassistant}, they often lack the scale and diversity needed to train models effectively for complex instruction-following tasks. Besides, even when data is available, the costs associated with finetuning on large-scale instruction datasets are prohibitively high, particularly for models with billions of parameters.

\subsection{Instruction Alignment via Model Merging}
\label{model_merge}
In contrast to multi-task learning, model merging is a training-free technique that directly fuses the weights of specialized LLMs to incorporate multiple capabilities without requiring access to the original training data~\cite{yang2024model}. Recent work has demonstrated that with a properly designed weight fusion scheme, model merging can achieve performance comparable to multi-task learning, making it an efficient method to equip LLMs with multiple capabilities~\cite{yu2024language}.

In literature, Model Soup represents a pioneering method in this direction, averaging the weights of different LLMs to create a single model that generalizes well across tasks~\cite{wortsman2022model}. Following this idea, Ilharco et al. proposed task arithmetic that averages the weight differences (i.e., task vectors) between input LLMs and their common base model; the resulting average is then added back to the base model to produce the merged model~\cite{ilharco2022editing}. Building on task arithmetic, TIES~\cite{yadav2024ties} enhances the approach by sparsifying the task vectors and incorporating a sign consensus algorithm prior to weight fusion. DELLA~\cite{deep2024della} further advances TIES via adaptively pruning less important weights with specific hyperparameters. 
However, all of these methods neglect the underlying geometric properties of LLM weights, which can lead to merged models with suboptimal performance.
In contrast, \name aims to merge a chip LLM with a well-instructed general LLM through geodesic interpolation. This geometric-aware technique allows us to smoothly blend the weights of the original LLMs along the shortest path 
on a Riemannian manifold. As a result, \name produces a merged model that effectively combines chip domain knowledge with instruction alignment, outperforming previous model merging techniques as detailed in Table \ref{openroad}. 
Notably, while \name has potential applications in other domains, this work primarily focuses on hardware-related QA tasks.

\section{The Proposed Approach}
\label{method}
\begin{figure}[t!]
\begin{center}
	\includegraphics[width=\columnwidth]{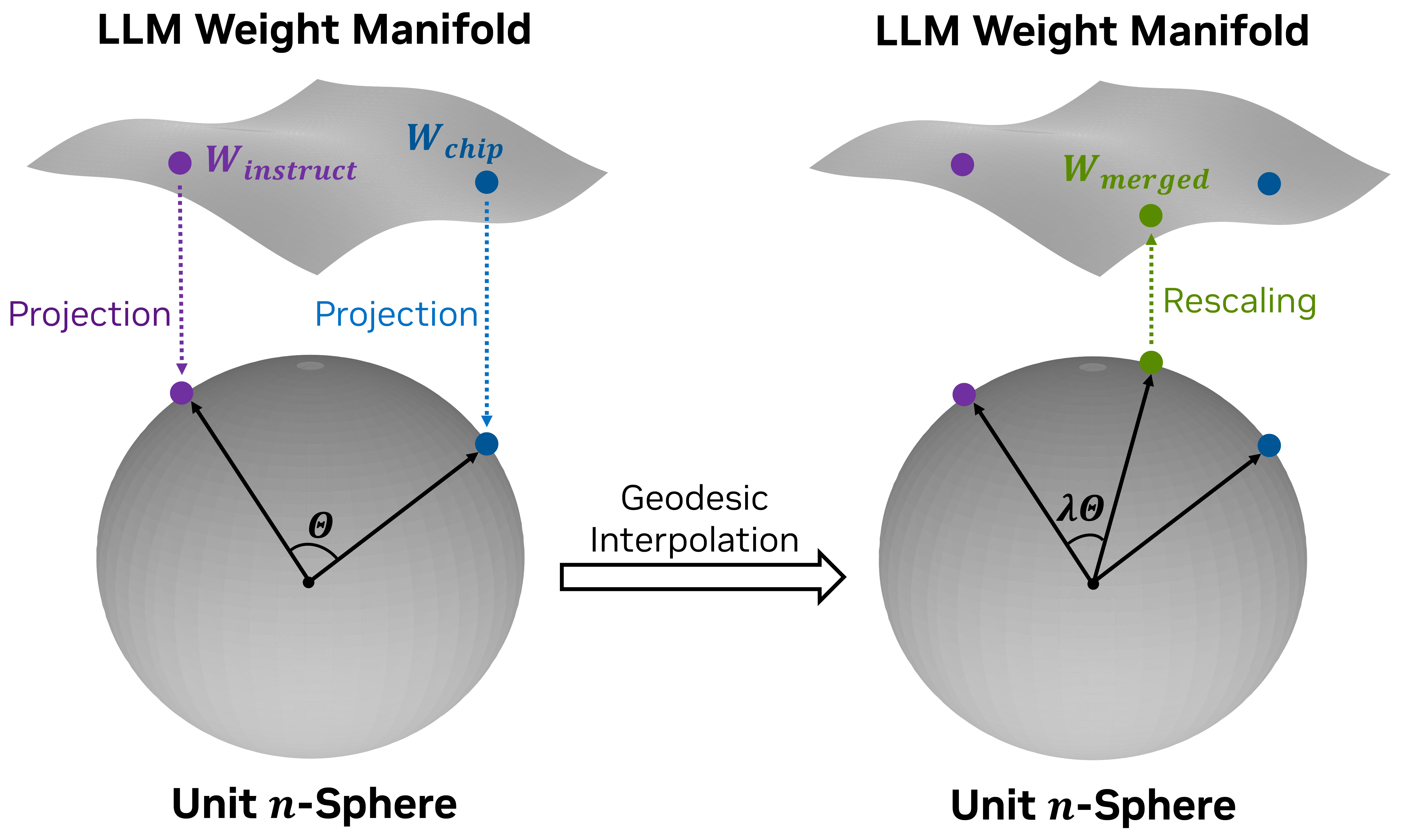}
	\caption{An overview of \name.} 
    \vspace{-18pt}
     \label{figure:overview}
\end{center}
\end{figure}
\textbf{Problem formulation} -- 
Let \( M_{\text{chip}} \) denote a chip LLM and \( M_{\text{instruct}} \) denote a general-purpose instruction LLM. 
Let $\{W^{(l)} | \, l=1,...,L\}$ denote the complete set of weights for an $L$-layer LLM, encompassing weights from the embedding layer, normalization layers, self-attention layers, and feed-forward layers.
For each layer \( l \), let \( W^{(l)}_{\text{chip}} \in \mathbb{R}^{p \times q} \) and \( W^{(l)}_{\text{instruct}} \in \mathbb{R}^{p \times q} \) represent the weight matrices of \( M_{\text{chip}} \) and \( M_{\text{instruct}} \), respectively. 
Our goal is to develop a merging function \( f \) such that: 
\[
W^{(l)}_{\text{merge}} = f(\, W^{(l)}_{\text{chip}}, \, W^{(l)}_{\text{instruct}} \,)
\]
The resulting \( W^{(l)}_{\text{merge}} \in \mathbb{R}^{p \times q} \) serves as the weights for the \( l \)-th layer in the merged model \( M_{\text{merge}} \), which aims to combine the strengths of both \( M_{\text{chip}} \) and \( M_{\text{instruct}} \). For brevity, we omit the layer index \( l \) in the following sections unless explicitly mentioned. Notably, our problem formulation implicitly assumes that the input models share the same architecture, meaning their respective weight matrices \( W_{\text{chip}} \) and \( W_{\text{instruct}} \) are conformable for merging. This assumption generally holds in practice; for example, ChipNeMo is trained based on LLaMA2-70B-Base, which has the same architecture as the instruction model LLaMA2-70B-Chat that is publicly available. 

Figure \ref{figure:overview} provides an overview of our proposed approach, \name. 
Next, we are going to first present the motivation for using geodesic interpolation in our merging method in Section \ref{geo_inter}. Then, Section \ref{slerp} introduces \name that computes geodesic interpolation along a unit $n$-sphere. Finally, we analyze the complexity of \name in Section \ref{complexity}.
It is worth noting that ChipAlign operates under the assumption that both chip and instruction LLMs are already available. Details on how to obtain these LLMs are provided in Section \ref{setup}.

\subsection{Model Merging via Geodesic Interpolation}
\label{geo_inter}
Neural network weights can be viewed as points on a high-dimensional Riemannian manifold~\cite{amari1992information,amari1994information}. 
This perspective opens the door for leveraging powerful geometric tools, such as geodesic interpolation, to transition smoothly between two sets of model weights. Geodesic interpolation follows the shortest path on the manifold between two points, providing a structured way to transition through the weight space without compromising on important model properties. Such interpolation techniques can be particularly useful for merging LLMs that specialize in different tasks. Given two points on the manifold corresponding to LLM weights optimized for distinct objectives, geodesic interpolation enables us to find a new point on the manifold that lies near both models, thereby inheriting strengths from each. This approach allows us to combine properties such as chip domain knowledge and instruction alignment into a single LLM.

However, a significant challenge arises due to the computational intractability of performing geodesic interpolation in high-dimensional manifolds. To overcome this, we project the model weights onto a specific type of Riemannian manifold, the unit $n$-sphere, where geodesic interpolation can be performed in linear time using a canonical form, as discussed in Section \ref{slerp}. This method provides an efficient way to merge model capabilities while respecting the geometric properties of the high-dimensional weight space.

\subsection{Geodesic Interpolation on Weight Manifold}
\label{slerp}
Our approach \name adopts geodesic interpolation between the weights of a chip LLM and an instruction-aligned LLM. 
To this end, we first formally define the unit $n$-sphere in the following.

\begin{definition}
\label{definition}
The \textit{unit \(n\)-sphere \(S^n\)} is the set of all points in \((n+1)\)-dimensional Euclidean space \(\mathbb{R}^{n+1}\) that are at a unit distance from the origin:
\[
S^n = \{ w \in \mathbb{R}^{n+1} : \|w\| = 1 \}
\]
where \(\|\cdot\|\) denotes the Euclidean norm.
\end{definition}

To efficiently perform geodesic interpolation between the weights of these models, \name\ first projects the weight matrices \( W_{\text{chip}} \) and \( W_{\text{instruct}} \) onto a unit $n$-sphere as follows: 
\[
\text{Norm}_{\text{chip}} = \|W_{\text{chip}}\|_F, \quad \text{Norm}_{\text{instruct}} = \|W_{\text{instruct}}\|_F
\]
\[
\bar{W}_{\text{chip}} = \frac{W_{\text{chip}}}{\text{Norm}_{\text{chip}}}, \quad \bar{W}_{\text{instruct}} = \frac{W_{\text{instruct}}}{\text{Norm}_{\text{instruct}}}
\]

where \( \| \cdot \|_F \) denotes the Frobenius norm. 
According to Definition \ref{definition}, \( \bar{W}_{\text{chip}} \in \mathbb{R}^{p \times q} \) and \( \bar{W}_{\text{instruct}} \in \mathbb{R}^{p \times q} \) now reside on a unit $n$-sphere ($n=p \times q-1$), the geodesic between them is represented by the arc connecting the two points on the sphere. This allows us to perform geodesic interpolation by interpolating along this arc, which can be efficiently achieved 
using the following lemma:
\begin{lemma}
\label{lemma}
Given \( \bar{W}_{\text{chip}} \) and \( \bar{W}_{\text{instruct}} \) lie on a unit $n$-sphere, the geodesic interpolation between them can be expressed as:
$$\bar{W}_{\text{merge}} = \frac{sin(\lambda \Theta)}{sin(\Theta)} \bar{W}_{\text{chip}} + \frac{sin((1-\lambda) \Theta)}{sin(\Theta)} \bar{W}_{\text{instruct}}$$
where $\Theta = arccos(\bar{W}_{\text{chip}}, \bar{W}_{\text{instruct}})$ is the angle between $\bar{W}_{\text{chip}}$ and $\bar{W}_{\text{instruct}}$, and $\lambda \in [0,1]$ determines the interpolation point along the geodesic, with \( \lambda = 0 \) corresponding to \( \bar{W}_{\text{instruct}} \) and \( \lambda = 1 \) corresponding to \( \bar{W}_{\text{chip}} \).
\end{lemma}

The proof for Lemma \ref{lemma} is available in~\cite{shoemake1985animating}. Lemma \ref{lemma} allows us to generate a continuum of models between the two input models along the unit $n$-sphere, with \( \lambda \) determining the degree to which each model’s weights are retained in the merged model. 

Finally, to restore the magnitude of the original weight matrices, we rescale the interpolated weights back to the manifold by applying the Frobenius norms:
\[
W_{\text{merge}} = \text{Norm}_{\text{chip}}^{\lambda} \cdot \text{Norm}_{\text{instruct}}^{1 - \lambda} \cdot \bar{W}_{\text{merge}}
\]
This process results in \( W_{\text{merge}} \), the weight matrix for each layer in the merged model \( M_{\text{merge}} \), effectively combining instruction alignment and chip design expertise in a single model. 

\begin{figure*}[t!]
\begin{center}
	\includegraphics[width=0.85\textwidth]{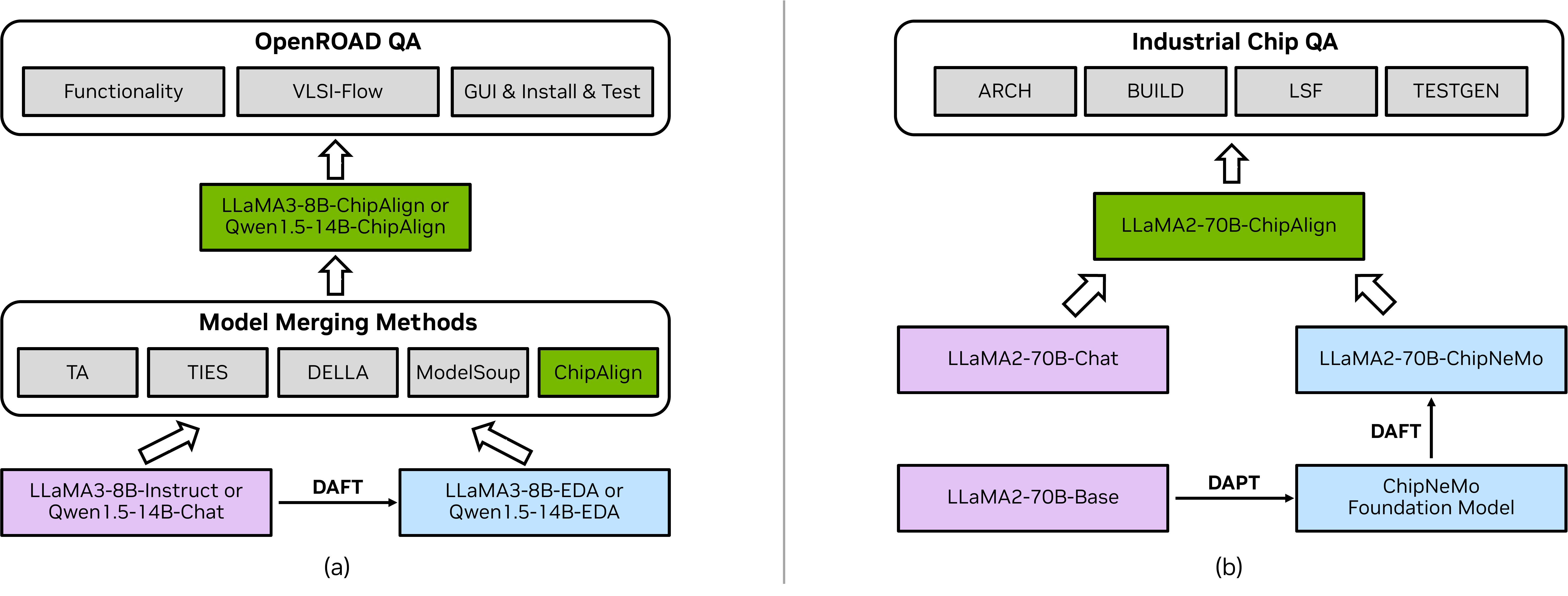}
	\caption{(a) An overview of developing merged models for the OpenROAD QA benchmark, featuring \name as the selected merging approach; (b) An overview of generating LLaMA2-70B-\name for industrial chip QA benchmarks.} 
    \vspace{-20pt}
        \label{figure:setup}
\end{center}
\end{figure*}
\textbf{Discussion} -- It is worth noting that an important advantage of \name is its ease of implementation, incorporating only a single hyperparameter $\lambda$. Our sensitivity analysis in Section \ref{sen_ana} reveals that $\lambda=0.6$ leads to peak performance across different LLM backbones. This minimizes the burden of hyperparameter tuning, making \name particularly easy to adopt for large-scale domain-adapted models. We further analyze the complexity of \name in Section \ref{complexity} to highlight its efficiency advantages.

\subsection{Complexity Analysis of \name}
\label{complexity}
\name projects the weight matrices of input LLMs, each with $n$ parameters, onto a unit $n$-sphere, requiring a single pass through the model weights in $O(n)$ time. It then performs geodesic interpolation along the arc between these weights and rescales the interpolated weights in $O(n)$ time. Consequently, the total time complexity for \name is $O(n)$. Similarly, its space complexity is $O(n)$, covering the storage of initial, projected, and final weights. This efficiency enables \name to handle billion-parameter scale LLMs with minimal computational overhead.

\section{Experiment}
We have conducted a comprehensive evaluation of \name, comparing its performance against state-of-the-art (SoTA) baselines on the OpenROAD QA benchmark and industrial production-level chip QA benchmarks. To further evaluate \name's capabilities, we assess its instruction alignment on the IFEval benchmark and its domain knowledge on multiple-choice QA benchmarks covering EDA script generation, bug summarization, and circuit design. Finally, we conduct a sensitivity analysis on the hyperparameter $\lambda$ in \name.
\subsection{Experimental Setup}
\label{setup}
\textbf{OpenROAD QA} -- As shown in Figure \ref{figure:setup}(a), we consider two well-aligned instruction LLMs that are publicly available: Qwen1.5-14B-Chat and LLaMA3-8B-Instruct. Following the methodology of Pu et al.\cite{pu2024customized} and Sharma et al.\cite{sharma2024openroad} to generate strong domain-adapted LLMs, we apply retrieval augmented DAFT to both models using around 2K context-query-answer training triplets from OpenROAD QA~\cite{zhang2024raft}. 
Specifically, we perform DAFT on each training QA pair along with its golden context, and adopt low-rank adaptation (LoRA) with a rank of $8$ and an alpha of $16$. We train the models over $20$ epochs with a learning rate of $2\times10^{-4}$ and a batch size of $1$. Both models are trained on four nodes of a computing cluster, each node being equipped with eight A100 GPUs, each with 80GB of memory.
This process results in the creation of Qwen1.5-14B-EDA and LLaMA3-8B-EDA.
Once both the instruction and domain-adapted LLMs are available, we use \name with $\lambda = 0.6$ to fuse their weights, producing Qwen1.5-14B-\name and LLaMA3-8B-\name. Remarkably, the fusion process takes only $10$ minutes on a CPU with $48$ cores running at $2.5$ GHz. Additionally, we compare \name with various popular model merging baselines such as task arithmetic (TA), TIES, DELLA, and Model Soup, adopting the recommended hyperparameters from their respective publications. Besides, we further compare \name against the general SoTA model GPT-4 Turbo and RAG-EDA, a Qwen1.5-14B-Chat based LLM highly customized for this benchmark~\cite{pu2024customized}. To quantitatively assess model performance, we follow Pu et al. to adopt the ROUGE-L scoring method for comparing the generated LLM responses with the golden answers~\cite{lin2004rouge}. We have found that the ROUGE-L score is a more representative metric on this benchmark than either the BLEU~\cite{papineni2002bleu} or UniEval~\cite{zhong2022towards} scores.


\textbf{Industrial chip QA} -- As indicated by Figure \ref{figure:setup}(b), we select what is possibly the largest chip LLM, LLaMA2-70B-ChipNeMo, as our baseline. This model is developed by first undergoing DAPT on 24 billion tokens derived from chip design documents and code, using the LLaMA2-70B-Base model, followed by DAFT on 57K instruction pairs. More training details are available in \cite{liu2023chipnemo}. Additionally, we choose LLaMA2-70B-Chat as the instruction model that is publicly available and has the same architecture as LLaMA2-70B-ChipNeMo. We set $\lambda=0.6$ in \name for merging LLaMA2-70B-ChipNeMo and LLaMA2-70B-Chat to produce LLaMA2-70B-\name. Owing to the linear complexity of \name, the weight fusion process takes 43 minutes on a CPU equipped with $48$ cores operating at $2.5$ GHz. Besides, we utilize a GPT-4 based grader to assess the quality of each LLM response by comparing it with the golden answer. The grader assigns scores in $\{0,25,50,75,100\}$, where a higher score indicates better answer quality.

\begin{figure*}[t!]
\begin{center}
	\includegraphics[width=\textwidth]{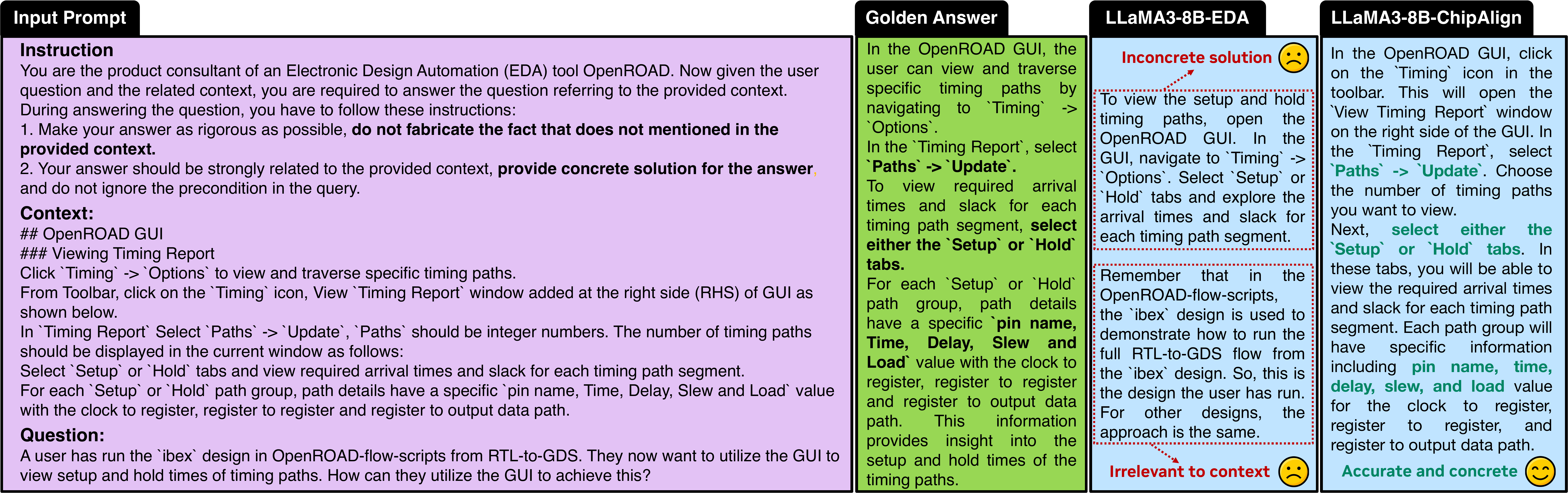}
	\caption{Comparison of model responses on the OpenROAD QA benchmark.} 
    \vspace{-10pt}
        \label{figure:prompt}
\end{center}
\end{figure*}
\begin{table*}[ht]
\centering
\captionsetup{width=\textwidth}
\caption{ROUGE-L scores on the OpenROAD QA benchmark --- $^*$ denotes the results sourced from~\cite{pu2024customized}.}
\label{openroad}
\begin{adjustbox}{width=\textwidth,center}
\begin{tabular}[t]{lcccccccc}
\toprule
 &  \multicolumn{4}{c}{\textbf{Golden Context}} & \multicolumn{4}{c}{\textbf{RAG Context}} \\
\cmidrule(r){2-5} \cmidrule(r){6-9}
\textbf{Method}   & Functionality   & VLSI Flow    & GUI \& Install \& Test   & All & Functionality   & VLSI Flow    & GUI \& Install \& Test   & All \\
\midrule
GPT-4 Turbo       &  $0.280$  &  $0.338$ & $0.367$  &  $0.315$ & $0.250$  &  $0.274$ & $0.331$  &  $0.276$ \\
RAG-EDA$^*$        &  $0.319$  &  $0.326$ & $0.374$  &  $0.334$ & $0.281$  &  $0.269$ & $0.302$  &  $0.283$ \\
\midrule
\midrule
Qwen1.5-14B-Chat        &  $0.276$  &  $0.318$ & $0.354$  &  $0.305$ & $0.244$  &  $0.264$ & $0.301$  &  $0.263$ \\
Qwen1.5-14B-EDA        &  $0.315$  &  $0.337$ & $0.385$  &  $0.338$ & $0.282$  &  $0.270$ & $0.335$  &  $0.292$ \\
\midrule
Qwen1.5-14B-TA        &  $0.316$  &  $0.346$ & $0.392$  &  $0.342$ & $0.285$  &  $0.288$ & $\mathbf{0.358}$  &  $0.303$ \\
Qwen1.5-14B-TIES        &  $0.300$  &  $0.352$ & $0.368$  &  $0.329$ & $0.255$  &  $0.279$ & $0.338$  &  $0.281$ \\
Qwen1.5-14B-DELLA        &  $0.295$  &  $0.343$ & $0.382$  &  $0.328$ & $0.254$  &  $0.278$ & $0.304$  &  $0.272$ \\
Qwen1.5-14B-ModelSoup        &  $0.317$  &  $0.359$ & $0.389$  &  $0.345$ & $0.295$  &  $0.278$ & $0.358$  &  $0.306$ \\
Qwen1.5-14B-\name        &  $\mathbf{0.354}$  &  $\mathbf{0.366}$ & $\mathbf{0.403}$  &  $\mathbf{0.369}$ & $\mathbf{0.305}$  &  $\mathbf{0.294}$ & $0.354$  &  $\mathbf{0.314}$ \\
\midrule
\midrule
LLaMA3-8B-Instruct        &  $0.277$  &  $0.307$ & $0.372$  &  $0.308$ & $0.258$  &  $0.232$ & $0.345$  &  $0.273$ \\
LLaMA3-8B-EDA        &  $0.322$  &  $0.360$ & $0.367$  &  $0.342$ & $0.283$  &  $0.265$ & $0.345$  &  $0.294$ \\
\midrule
LLaMA3-8B-TA        &  $0.284$  &  $0.362$ & $0.372$  &  $0.325$ & $0.267$  &  $0.292$ & $0.344$  &  $0.292$ \\
LLaMA3-8B-TIES        &  $0.286$  &  $0.332$ & $0.376$  &  $0.319$ & $0.256$  &  $0.281$ & $0.347$  &  $0.285$ \\
LLaMA3-8B-DELLA        &  $0.289$  &  $0.340$ & $0.365$  &  $0.320$ & $0.263$  &  $0.275$ & $0.353$  &  $0.288$ \\
LLaMA3-8B-ModelSoup        &  $0.333$  &  $0.370$ & $\mathbf{0.429}$  &  $0.365$ & $0.288$  &  $0.286$ & $0.368$  &  $0.307$ \\
LLaMA3-8B-\name        &  $\mathbf{0.362}$  &  $\mathbf{0.385}$ & $0.427$  &  $\mathbf{0.383}$ & $\mathbf{0.304}$  &  $\mathbf{0.300}$ & $\mathbf{0.392}$  &  $\mathbf{0.325}$ \\
\bottomrule
\end{tabular}
\end{adjustbox}
\vspace{-15pt}

\end{table*}

\subsection{Evaluation on OpenROAD QA}
We evaluate all models using 90 high-quality context-query-answer triplets from the OpenROAD QA benchmark, which all follow the same instruction shown in Figure \ref{figure:prompt}. For each triplet, the context is derived either from the golden context corresponding to the given query or from the retrieved context via retrieval-augmented generation (RAG) using OpenROAD documentation. To establish a strong RAG pipeline, we employ the \textit{bge-large-en-v1.5} for text embedding, \textit{BM25} for lexical retrieval, and \textit{bge-reranker-large} for re-ranking~\cite{xiao2023c}. For the purpose of reproducibility, we set the temperature parameter to 
$0.0$ for all models during response generation.

Table \ref{openroad} showcases that \name surpasses all existing model merging methods in most cases, achieving a ROUGE-L score improvement of up to $6.4\%$. This empirical evidence confirms that our geometric-aware method produces a superior merged model compared to other baselines. Moreover, the merged models generated by \name consistently outperform their corresponding EDA models across different LLM backbones, achieving performance improvements of up to $3.9\%$—a nontrivial margin for this benchmark. This improvement is attributed to \name's efficacy in enhancing instruction alignment of domain-adapted models, resulting in responses that better align with the provided instructions, as evidenced in Figure \ref{figure:prompt} and Table \ref{ifeval}.

\subsection{Evaluation on Industrial Chip QA}
\begin{figure*}[t!]
\begin{center}
	\includegraphics[width=\textwidth]{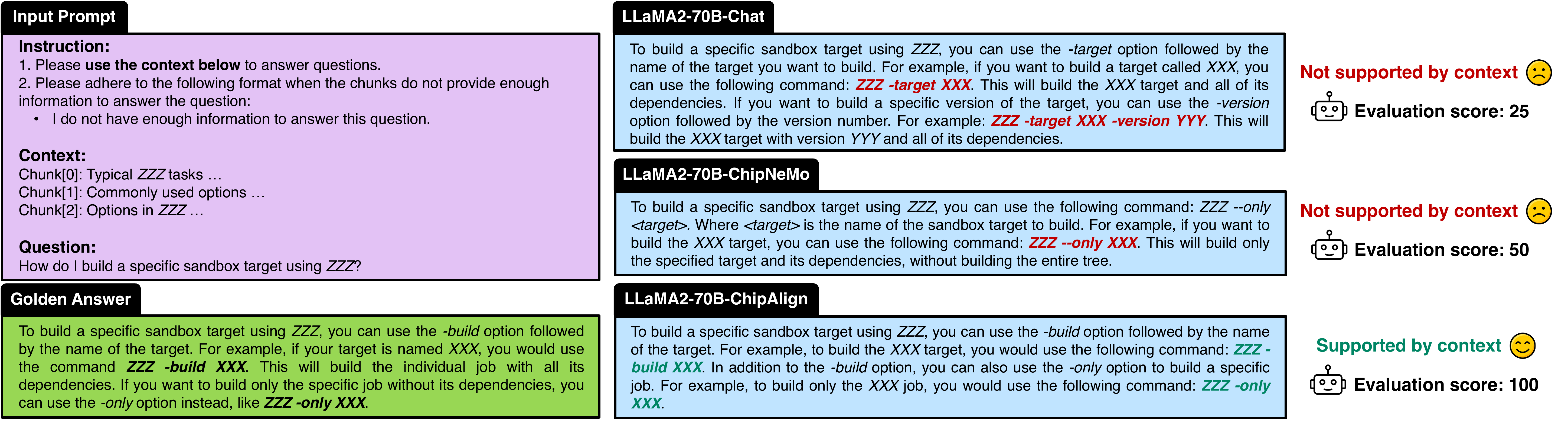}
	\caption{An example from the BUILD category of the industrial chip QA benchmark --- Due to space limitations, only the first few words of each context chunk are shown, omitting full context. Besides, we redact the build tool and target names using \textit{ZZZ} and \textit{XXX}, respectively.} 
    \vspace{-10pt}
        \label{figure:multiturn}
\end{center}
\end{figure*}
\begin{table*}[ht]
\centering
\captionsetup{width=\textwidth}
\caption{GPT4-aided evaluation scores on industrial production-level chip QA benchmarks.}
\label{chip_chat}
\begin{adjustbox}{width=\textwidth,center}
\begin{tabular}[t]{lcccccccccc}
\toprule
 &  \multicolumn{5}{c}{\textbf{Single Turn}} & \multicolumn{5}{c}{\textbf{Multi Turn}} \\
\cmidrule(r){2-6} \cmidrule(r){7-11}
\textbf{Method}   & ARCH   & BUILD    & LSF   & TESTGEN & All   & ARCH   & BUILD    & LSF   & TESTGEN & All \\
\midrule
LLaMA2-70B-Chat       &  $40.00$  &  $70.50$ & $54.25$  &  $66.75$ & $57.00$  &  $10.00$ & $41.00$  &  $10.50$ & $50.00$  &  $30.25$ \\
LLaMA2-70B-ChipNeMo        &  $45.00$  &  $81.75$ & $60.50$  &  $87.50$ & $66.75$  &  $40.00$ & $\mathbf{81.75}$  &  $52.00$ & $33.25$  &  $54.50$ \\
LLaMA2-70B-\name        &  $\mathbf{57.50}$  &  $\mathbf{84.00}$ & $\mathbf{70.75}$  &  $\mathbf{91.75}$ & $\mathbf{74.25}$  &  $\mathbf{42.50}$ & $75.00$  &  $\mathbf{60.50}$ & $\mathbf{79.25}$  &  $\mathbf{62.75}$ \\
\bottomrule
\end{tabular}
\end{adjustbox}
\vspace{-10pt}

\end{table*}

We have extended our evaluation of \name to include industrial production-level chip QA benchmarks, consisting of 39 practical questions from hardware design engineers across domains such as hardware architecture (ARCH), build processes (BUILD), job scheduling (LSF), and verification (TESTGEN). In addition to a single-turn setting where each question is treated independently, we also explore a multi-turn setting to simulate real-world scenarios where engineers may pose follow-up questions based on previous interactions. As illustrated in Figure \ref{figure:multiturn}, each input prompt includes a question, relevant contexts obtained through RAG, and multiple instructions, necessitating strong instruction alignment in LLMs. We set the temperature parameter to $0.0$ for all model responses.

Table \ref{chip_chat} showcases that the merged model LLaMA2-70B-\name consistently outperforms both source models (LLaMA2-70B-Chat and LLaMA2-70B-ChipNeMo) by a margin of up to $8.25\%$. Besides, Figure \ref{figure:multiturn} illustrates that 
both LLaMA2-70B-Chat and LLaMA2-70B-ChipNeMo fail to follow the instructions, which mandate a model to answer questions using information from the relevant context.
In contrast, LLaMA2-70B-\name strictly adheres to the provided instructions and generate correct answers supported by the context. This confirms \name's efficacy with the largest existing chip LLMs, which is further analyzed in Section \ref{instruct_domain}.

\subsection{Evaluation on Instruction Alignment and Domain Knowledge}
\label{instruct_domain}
\begin{table}[ht]
\centering
\captionsetup{width=\textwidth}
\caption{Instruction-following accuracy ($\%$) on IFEval.}
\label{ifeval}
\begin{adjustbox}{width=\columnwidth,center}
\begin{tabular}[t]{lcccc}
\toprule
 &  \multicolumn{2}{c}{\textbf{Prompt Level}} & \multicolumn{2}{c}{\textbf{Instruction Level}} \\
\cmidrule(r){2-3} \cmidrule(r){4-5}
\textbf{Method}   & Strict   & Loose    & Strict   & Loose \\
\midrule
LLaMA3-8B-Instruct       &  $67.5$  &  $\mathbf{75.4}$ & $76.1$  &  $\mathbf{82.6}$ \\
LLaMA3-8B-EDA        &  $60.0$  &  $64.6$ & $69.6$  &  $74.1$ \\
LLaMA3-8B-\name        &  $\mathbf{69.9}$  &  $74.1$ & $\mathbf{78.3}$  &  $81.9$ \\
\midrule
LLaMA2-70B-Chat       &  $43.2$  &  $50.1$ & $55.2$  &  $61.5$ \\
LLaMA2-70B-ChipNeMo        &  $27.9$  &  $32.0$ & $39.8$  &  $44.4$ \\
LLaMA2-70B-\name        &  $\mathbf{54.3}$  &  $\mathbf{61.0}$ & $\mathbf{65.0}$  &  $\mathbf{70.3}$ \\
\bottomrule
\end{tabular}
\end{adjustbox}

\end{table}

\begin{figure}[ht]
\vspace{-10pt}
\begin{center}
	\includegraphics[width=0.9\columnwidth]{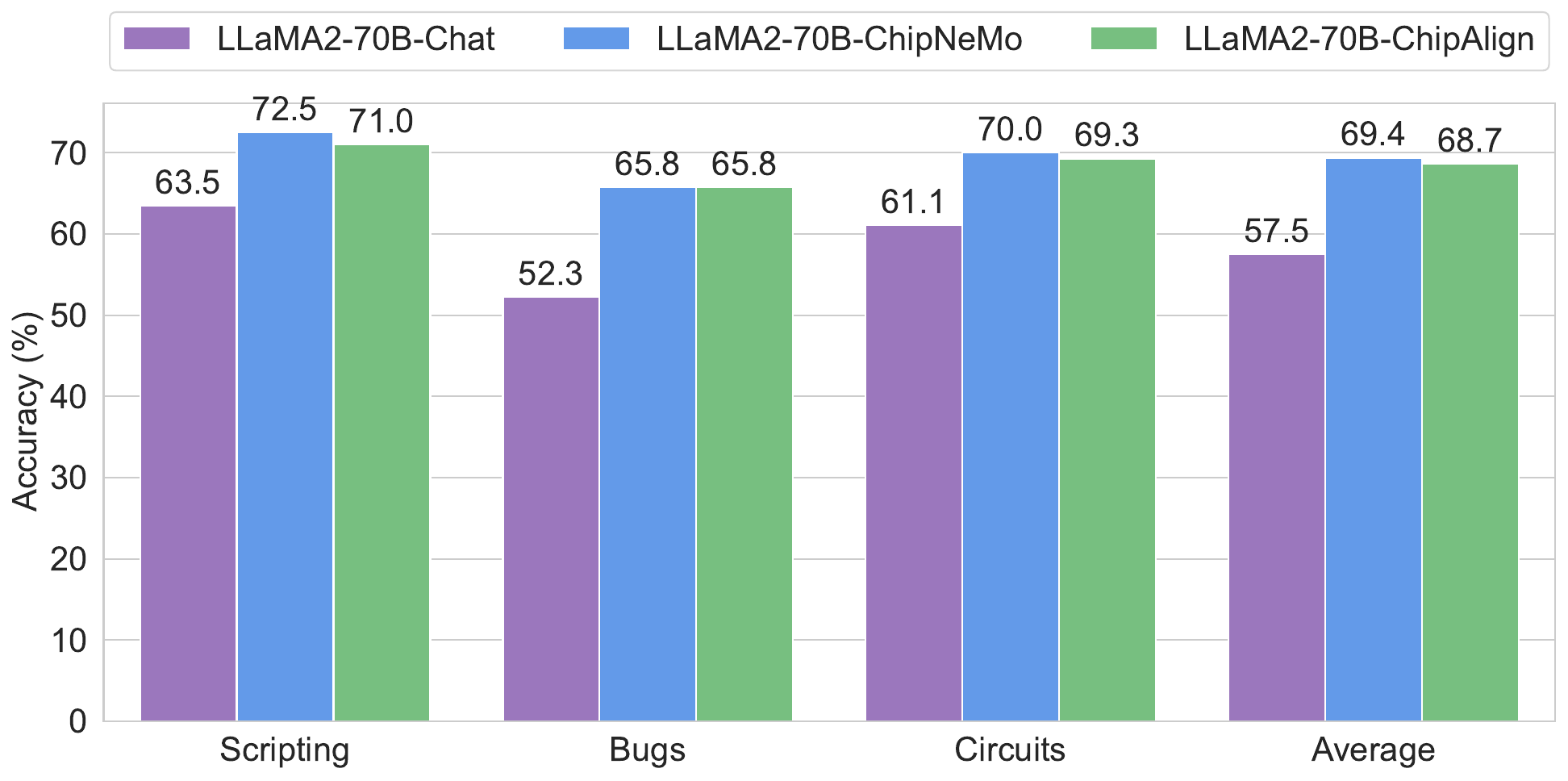}
	\caption{Results on multi-choice chip QA benchmark.}
    \vspace{-20pt}
     \label{figure:chip_qa}
\end{center}
\end{figure}
In addition to evaluating \name on both the OpenROAD QA and industrial chip QA benchmarks, which demand a joint capability of instruction alignment and domain knowledge, we further assess these two capabilities of \name separately.

Specifically, we assess instruction alignment using IFEval, a well-known benchmark designed to test LLMs' ability to follow instructions. IFEval comprises 541 prompts featuring various general instructions, such as ``List exactly 10 possible names ...'' and ``Think step-by-step and then give your answer ...''. Accuracy for each LLM is calculated based on the number of prompts or instructions that are either strictly (Strict Accuracy) or loosely (Loose Accuracy) adhered to. 

Our results presented in Table \ref{ifeval} showcase that LLaMA3-8B-\name significantly outperforms LLaMA3-8B-EDA in instruction-following accuracy and matches the performance of LLaMA3-8B-Instruct. More importantly, LLaMA2-70B-\name not only improves the accuracy of LLaMA2-70B-ChipNeMo by $26.6\%$ on average, but also considerably surpasses the source instruction model, LLaMA2-70B-Chat. This enhanced performance can be attributed to the integration of the OASST instruction dataset~\cite{kopf2024openassistant} and the SteerLM alignment strategy~\cite{dong2023steerlm} during the training of LLaMA2-70B-ChipNeMo, which imbued its weights with decent instructional knowledge complementary to that of LLaMA2-70B-Chat. Consequently, \name benefits from the combined instructional knowledge of LLaMA2-70B-ChipNeMo and LLaMA2-70B-Chat through the process of weight fusion. This fusion results in a merged model with stronger instruction alignment capabilities compared to both source models.

For evaluating pure chip domain knowledge, we utilize multi-choice chip QA benchmarks from~\cite{liu2023chipnemo} that contain no instructions. Figure \ref{figure:chip_qa} shows that \name performs on par with ChipNeMo across the domains of EDA scripts, bugs, and circuits, highlighting its ability to preserve domain knowledge after weight fusion. Furthermore, the comparative performance of LLaMA2-70B-Chat, LLaMA2-70B-ChipNeMo, and LLaMA2-70B-\name is visualized in Figure \ref{figure:radar}, providing a comprehensive overview of their respective capabilities.

\subsection{Sensitivity Analysis on \texorpdfstring{$\lambda$}{}}
\label{sen_ana}
\vspace{-10pt}
\begin{figure}[ht]
\begin{center}
	\includegraphics[width=0.8\columnwidth]{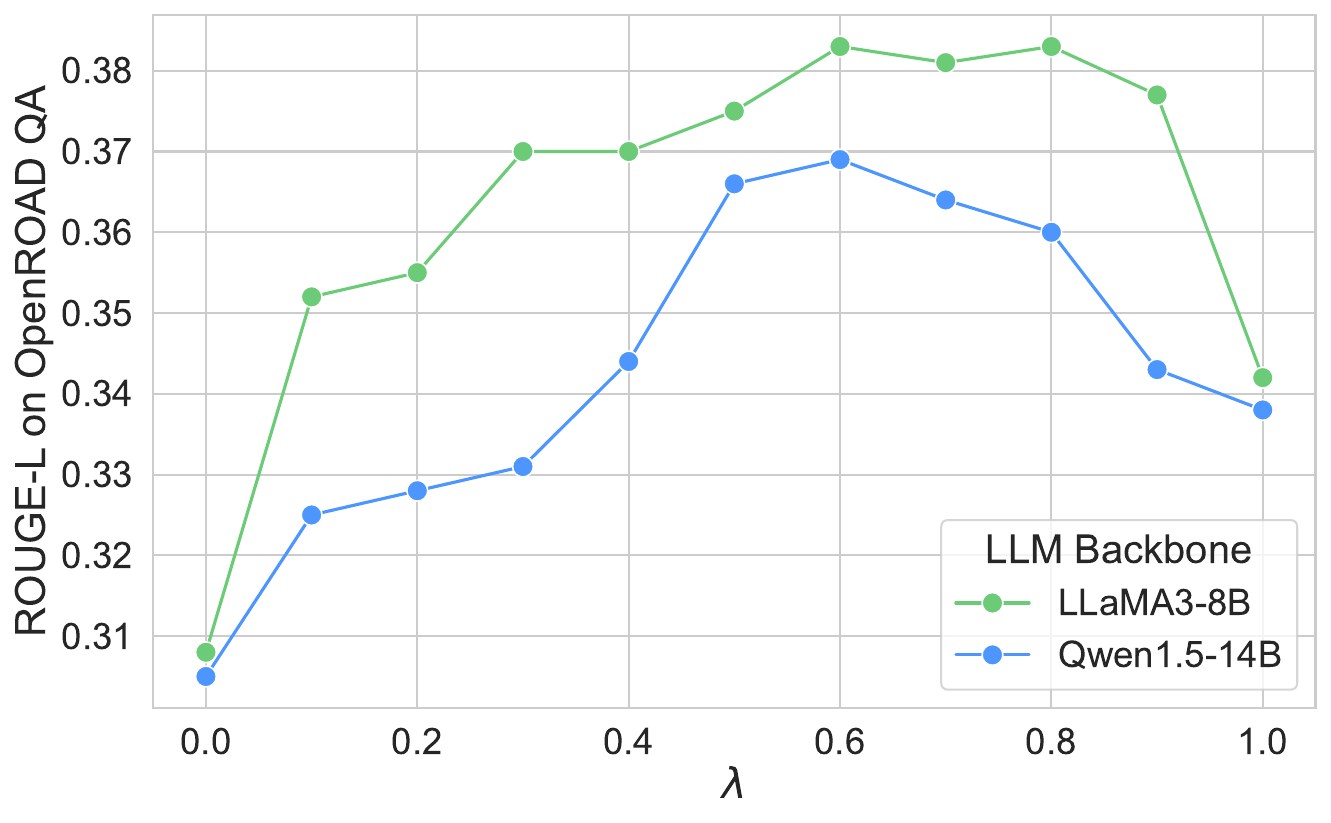}
    \vspace{-5pt}
	\caption{Sensitivity analysis on $\lambda$ in \name.} 
    \vspace{-10pt}
     \label{figure:analyze}
\end{center}
\end{figure}
As discussed in Section \ref{method}, \name involves a single hyperparameter $\lambda$ that determines how closely the merged model aligns with either the instruction model $M_{\text{instruct}}$ or the chip model $M_{\text{chip}}$ along the geodesic. We conduct a sensitivity analysis of $\lambda$ using both LLaMA3-8B and Qwen1.5-14B backbones on the OpenROAD QA benchmark. Notably, $\lambda=0$ and $\lambda=1$ correspond to the $M_{\text{instruct}}$ and $M_{\text{chip}}$, respectively. As depicted in Figure \ref{figure:analyze}, model performance initially increases rapidly from $M_{\text{instruct}}$ (i.e., the leftmost point), peaks at $\lambda=0.6$, and subsequently declines towards the performance level of $M_{\text{chip}}$ (i.e., the rightmost point). Hence, we recommend setting $\lambda=0.6$ as the default value in \name for practical applications.

\section{Conclusion}
This work introduces \name, a geometric-aware model merging approach to enhance instruction alignment in chip LLMs. \name treats the weights of a chip LLM and a general instruction LLM as two points on a Riemannian manifold and employs geodesic interpolation to create the merged model. Our results demonstrate that this merged model significantly improves instruction alignment in chip LLMs, yielding superior performance across various chip QA benchmarks.


\bibliographystyle{plain}
\bibliography{ref}

\end{document}